\documentclass[twocolumn,showpacs,preprintnumbers,amsmath,amssymb]{revtex4}
\usepackage{graphicx}% Include figure files
\usepackage{dcolumn}% Align table columns on decimal point
\usepackage{bm}% bold math
\usepackage[latin1]{inputenc}
\begin{document}

\title{Quantum q-breathers in a finite Bose-Hubbard chain: The case of two
interacting bosons}
\author{Jean Pierre Nguenang$^{1,2}$}
\author{R. A. Pinto$^{1}$}
\author{Sergej Flach$^{1}$}
\affiliation{1\ Max-Planck-Institut f\"ur Physik komplexer Systeme\\
N\"othnitzer Str. 38, 01187 Dresden, Germany}
\affiliation{2\ Fundamental physics laboratory: Group of Nonlinear physics and
Complex Systems.\\ Department of Physics. University of Douala\\
 P.O.Box:24157, Douala-Cameroon }

\date{\today}

\begin{abstract}
We study the spectrum and eigenstates of the quantum discrete
Bose-Hubbard Hamiltonian in a finite one-dimensional
lattice containing two bosons. The interaction between the bosons leads to an
algebraic
localization of the modified extended states in the normal mode space of the
noninteracting system. Weight
functions of the eigenstates in the space of normal modes are computed by using
numerical
diagonalization and perturbation theory. We find that staggered states do not
compactify in the dilute limit for large chains.

\end{abstract}


\maketitle

\section{Introduction}

The study of discrete breathers in
different physical systems has had
remarkable developments during the last two decades
\cite{FlachPhysRep295,physicstoday,Sievers,AubryPhysicaD103}. 
These
excitations are generic time-periodic and spatially localized solutions of the
underlying classical Hamiltonian lattice with translational invariance. 
Their spatial profiles localize exponentially for short-range interaction.
Recent
experimental observations of breathers in various systems
include such different cases as bond excitations in molecules, 
lattice vibrations and spin excitations in solids, electronic currents in
coupled Josephson junctions, light propagation in interacting optical
waveguides, cantilever vibrations in micromechanical arrays, cold atom
dynamics in Bose-Einstein condensates loaded on optical lattices, among others
\cite{SchwarzPRL83,SatoNature,SwansonPRL82,TriasPRL84,
BinderPRL84,EisenbergPRL81,FleischerNature422,SatoPRL90,EiermannPRL92}. In many
cases quantum
dynamics is important. Quantum breathers consist of
superpositions of nearly degenerate many-quanta bound states, with very long
times to tunnel from one lattice site to another
\cite{Fleurov,ScottPhysLettA119,BernsteinNonlin3,BernsteinPhysicaD68,WrightPhysicaD69,
Wang,Aubry,Flach1,AubryPhysicaD103,Fleurov1998,kalosakas2,Dorignac2004,Eilbeck2004,
Pinto,Schulman,Schulman2006,Proville2006}. 
Remarkably quantum breathers, though being extended states in a translationally
invariant
system, are characterized by exponentially localized weight functions, in
full
analogy to their classical counterparts.

Recently the application of these ideas to normal mode space allowed to
explain many facettes of the Fermi-Pasta-Ulam (FPU)
paradox \cite{Fermi}, which consists of the nonequipartition of energy among
the linear normal modes in a nonlinear chain. There the energy is localized
around
the initial normal mode which is excited. Introducing
the notion of q-breathers \cite{FlachPRL95,IvanchenkoPRL2006,FlachPRE2006}, 
which are time-periodic excitations 
localized in the
normal mode space, the FPU paradox and some related problems
were successfully explained. Despite the fact that the interaction in normal
mode space is long-ranged, it is selective and purely nonlinear, thus
q-breathers localize exponentially in normal mode space.

In this paper we
address the properties of quantum q-breathers. We study a one-dimensional
quantum lattice problem with two quanta.
By defining an appropriate weight function in the normal mode space we
explore the localization properties of the eigenstates of the system. We
observe localization of the weight function as a function of the wave
number, which we interprete as a signature of quantum q-breather
excitations. By using a numerical diagonalization of the Hamiltonian and
nondegenerate perturbation theory we find algebraic decay of the weight
function in the normal mode space, at variance to the exponential decay found
for 
q-breathers in the case of a classical nonlinear system. Another intriguing 
difference is based on the interference effects of two interacting quanta.
For the general case the quantum q-breather states approach 
the noninteracting eigenstates in the dilute limit of large
chains. However, states with Bloch momentum close to $\pm \pi$ 
keep their finite localization in that limit.

In section II we describe the model and introduce the basis we use to write
down the Hamiltonian matrix. In section III we review results on
the properties of two-quanta bound states - the simplest versions
of a quantum breather.
In section IV we
consider the case of extended states.
We introduce a weight function to describe
localization in the normal mode space, and obtain analytical results using
perturbation theory. We present our numerical results obtained
by diagonalization of the Hamiltonian matrix, comparing them to analytical
estimations. We conclude in section V.

\section{The model}

We study a one-dimensional periodic lattice with
$f$ sites described by the Bose-Hubbard (BH) model. This is a quantum version
of the discrete nonlinear Schr\"odinger equation, which has been used to
describe a
great variety of systems \cite{Scott1}. The BH Hamiltonian is given by
\cite{Eilbeck94}
\begin{equation}\label{eq:hamiltonian}
\hat{H} = \hat{H}_0 + \gamma\hat{H}_1,
\end{equation}
where
\begin{equation}
\hat{H}_0=-\sum_{j=1}^f b_j^+(b_{j-1}+b_{j+1}),
\end{equation}
\begin{equation}
\hat{H}_1 = -\frac{1}{2}\sum_{j=1}^f b_j^+b_j^+b_jb_j.   
\end{equation}
Here $ b_j^+$ and $b_j$ are the bosonic creation and annihilation operators
which
satisfy the commutation relations $[b_i,b_j^{+}]=\delta_{ij}$,
$[b_i^{+},b_j^{+}]=[b_i,b_j]=0$.
$\gamma$ is the parameter controlling the strength of
the interaction, and 
the chain of length $f$ is subject to periodic boundary conditions. The chain
is
translational invariant and the Hamiltonian (\ref{eq:hamiltonian})
commutes with the number operator $\hat{N}=\sum_{j=1}^f{b_j^+b_j}$, whose
eigenvalue is denoted by $n$. We consider the simplest  non-trivial case of $n=2$.
It is of direct relevance to studies and observations of 
bound two-vibron states \cite{Cohen1969,Kimball,Richter1988,GuyotSionnest1991,Dai1994,Chin1995,Jakob1996,JakobPr75,Pouthier2003JCP,Okuyama2001,Pouthier2003PRE,Edler2004}
 
In order to describe the quantum states, we use a number
state basis \cite{Eilbeck94} $|\Phi_n\rangle=|n_1,n_2,...,n_f\rangle$,  where
$n_i$ represents the number of bosons at site i
$(n=\sum{n_i})$. As an example $|0200000\rangle$ corresponds to a state with two bosons
on the second
site and zero bosons elsewhere. For a given number of bosons each eigenstate is
a linear combination of number states with fixed $n$.
In addition to the number of quanta $n$ there are $n-1$ further quantum
numbers which define the relative distance between the bosons.
For $n=2$ that reduces to defining one further relative distance $j-1$ between
the two quanta, which can take $(f+1)/2$ different values
in our case:
\begin{equation}
|\Psi_2\rangle=\sum_{j=1}^{\frac{f+1}{2}}v_j|\Phi_2^j\rangle.
\end{equation}
Due to translation invariance the eigenstates of $\hat{H}$
are also eigenstates of the translational operator $\hat{T}$,
where $\tau=\exp(ik)$ is its
eigenvalue with
$k=2\pi\nu/f$ being the Bloch wave number and $\nu\in [-(f-1)/2,
(f-1)/2]$. Due to periodic boundary conditions
$\hat{T}|n_1,n_2,\cdots,n_f\rangle=|n_f,n_1,n_2,\cdots,n_{f-1}\rangle$. For
the sake of simplicity we deal with an odd number of sites $f$.
Thus we can construct number states which are also Bloch states:
\begin{equation}
|\Phi_2^j\rangle=\frac{1}{\sqrt{f}}
\sum_{s=1}^f\Big(\frac{\hat{T}}{\tau}\Big)^{s-1}|1\underbrace{0\cdots0}_{j-1}1\cdots\rangle
.
\end{equation}
With
this basis we can derive the eigenenergies 
for each given Bloch wave number $k$ from
$\hat{H}_k|\Psi_n\rangle=E|\Psi_n\rangle$ after 
computing the eigenvalues of the matrix
with the same structure as in \cite{Eilbeck94} for the case of the BH
system:
\begin{eqnarray}
\label{matrix}
\hat{H}_k = -\left( \begin{array}{cccccc}
\gamma & q\sqrt{2} & & & & \\
q^*\sqrt{2} & 0 & q & & & \\
      & q^* & 0 & q & & \\
  &  & \ddots & \ddots &\ddots & \\
  &  & & q^* & 0 & q \\
  &  & &  & q^* & p
\end{array} \right),
\end{eqnarray}
with $q=1+\tau$ and $p=\tau^{-(f+1)/2} +
\tau^{-(f-1)/2}$. 
By varying the Bloch wavenumber in its irreducible range,
we obtain the eigenenergy spectrum shown in Fig.\ref{spectrum}.

\section{Bound states: Localization in real space}

\begin{figure}[h!]
\includegraphics[width=2.in]{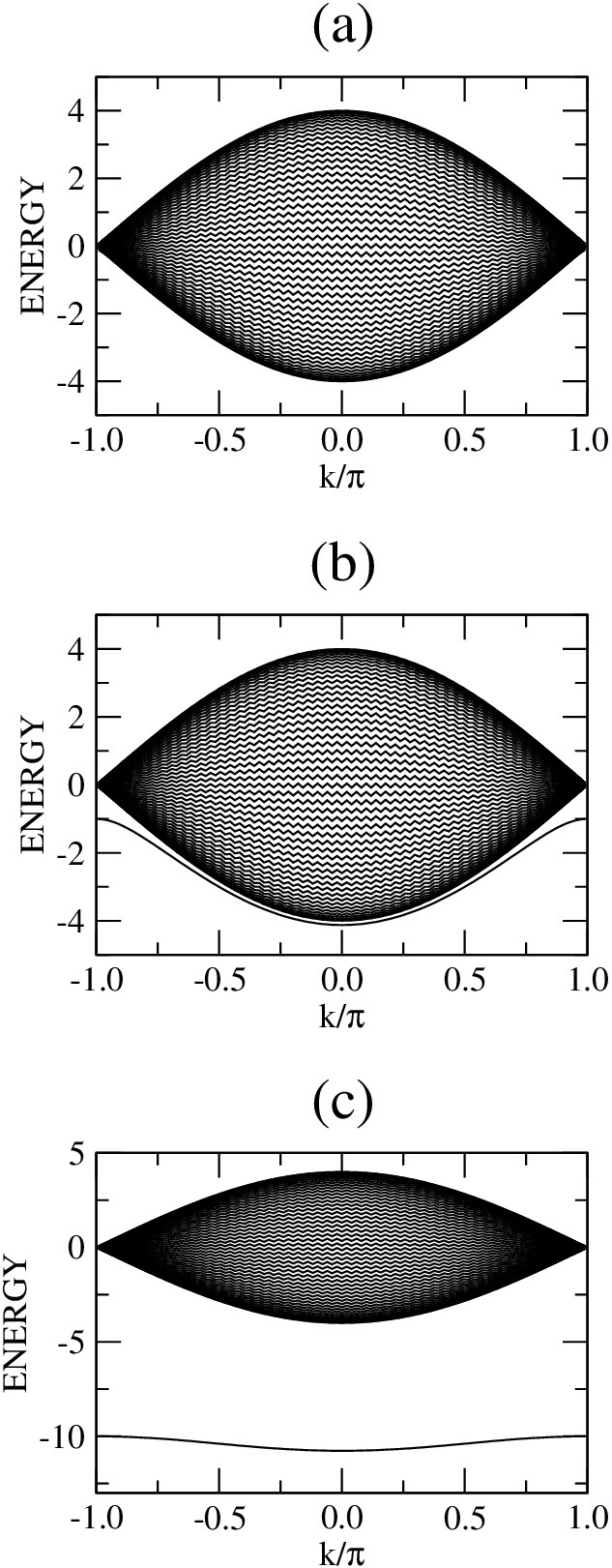}
\caption{\label{spectrum}Energy spectrum of the Bose-Hubbard model for
different values of the interaction $\gamma$: (a) $\gamma=0.1$, (b)
$\gamma=1.0$,
and (c) $\gamma=10$. Here $f=101$.}
\end{figure}

In Fig.\ref{spectrum}(a-c) we show that as the interaction parameter is
increasing, an isolated ground state eigenvalue $E_2(k)$ appears for each $k$
that
corresponds to a bound state \cite{Eilbeck94}. 
For this isolated ground state there is a high
probability of finding two quanta on the same
site. In the limit $f\to\infty$ the bound state eigenvalue has the analytical
expression \cite{Eilbeck94,Eilbeck03}
%{\bf Sergej, please add the reference of the book
%where the formulas for the
%eigenvector appear; the one you have where Eilbeck shows these results}
:
\begin{equation}
E_2(k) = -\sqrt{\gamma^2 + 16\cos^2{k/2}},
\end{equation}
and the corresponding (unnormalized) eigenvector $\mathbf{v}=(v_1,v_2,\ldots)$
is \cite{Eilbeck03}
\begin{equation}
\mathbf{v} = \left( \frac{1}{\sqrt{2}},\mu,\mu^2,\mu^3,\ldots \right),
\end{equation}
where
\begin{equation}
\mu = -\frac{(\gamma + E_2(k))e^{ik/2}}{4\cos(k/2)}.
\end{equation}
%We can prove that $|\mu|^2<1$ for $\gamma >0$.
A suitable weight function 
of this isolated ground state has the form:
\begin{equation}
C_{j} \equiv |v_j|^2 = |\mu|^{2(j-1)} = e^{2\lambda (j-1)}, \; \; j>1,
\end{equation}
where $C_1=1/2$ and $\lambda =\ln|\mu|$, Since $|\mu|^2<1$ for $\gamma \neq 0$,
the weight function shows exponential decay when the
distance between the two bosons increases. 
That result corresponds to the exponential localization of classical discrete breathers
\cite{FlachPhysRep295,physicstoday,Sievers,AubryPhysicaD103}.
However note that for $|k| \rightarrow \pi$ we have $\mu \rightarrow 0$
independently on the value of $\gamma \neq 0$.
Thus one obtains compact localization.
Note that it is said that a state is compact in a certain basis, if it occupies
a certain subspace, but has exactly zero overlap with the rest.

The compact localization for $|k| \rightarrow \pi$ is not observed in the
classical limit, and relies on the fact that the Schr\"odinger equation is a linear wave equation
which admits (destructive) interference effects.

%When $\gamma\gg\cos(k/2)$ we have:
%\begin{equation}
%\mu \approx \frac{2}{\gamma}\cos(k/2)e^{ik/2}.
%\end{equation}
%Thus for $|k| \rightarrow \pi$ we obtain compact localization $\mu -\rightarrow 0$
%independently on the value of $\gamma \neq 0$.
%%%%%%%%%%%%%%%%%%%%%%%%%%%%%%%%%%%%%%%%%%%%%%%%%%%%%%%%%%%%%%%%%%%%%%%%%
%%%%%%%%%%%%%%%%%%%%%%%%%%%%%%%%%%%%%%%%%%%%%%%%%%%%%%%%%%%%%%%%%%%%%%%%%%%%%%

\section{Quantum q-breathers: localization in normal mode space }

All the other states (except the bound state) form the two quanta
continuum. Their energies for $\gamma=0$ correspond to the sum of two single
particle energies with the constraint that the sum of their momenta equals the
Bloch momentum $k$. One arrives at 
\begin{equation}
E_{k,k_1}^0 = -2[\cos(k_1)+\cos(k_1+k)],
\end{equation}
where $k_1=\pi\nu_1/[(f+1)/2]-k/2$ is the conjugated momentum of the relative
coordinate
(distance) of both quanta and $\nu_1 = 1,\ldots,(f+1)/2$. $E_{k,k_1}^0$ has a
finite spread at fixed $k$ (see
Fig.\ref{spectrum}). However for $k=\pm\pi$ the spectrum becomes
degenerate. Thus for $|k\pm\pi|\ll 1$ the eigenenergies are very close (almost
degenerate). Remarkably the bounds of the spectrum for $\gamma\neq 0$ are very
well described by the $\gamma=0$ result. Increasing $\gamma$ at fixed $k$, the
eigenenergies will slightly move, but never cross. Thus a continuation of an
eigenstate at $\gamma=0$ to $\gamma\neq 0$ will preserve its relative ordering
with respect to the other eigenenergies.

For $\gamma\neq 0$ these quantum q-breather states will be deformed. In analogy
to
the study of the fate of normal modes in classical nonlinear systems
\cite{FlachPRL95,IvanchenkoPRL2006,FlachPRE2006}, we will study the changes of
the two-quanta continuum. For finite $f$ and $\gamma$ the new states will be
spread in the basis of the $\gamma=0$ continuum. For $f\to\infty$ one expects
that the new states become again identical with the $\gamma=0$ states, since
the two quanta will meet on the lattice with less probability as $f$
increases. Thus we will test the  compactification of the new states in the
$\gamma=0$ eigenstate basis both for $\gamma\to 0$ and for $f\to\infty$.

We compute the weight functions in normal
mode space in order to probe the signature of quantum q-breathers. For this
purpose we start by using perturbation theory to set up
these weight functions, where $H_1$ is the perturbation. We fix the Bloch
momentum $k$, and choose an eigenstate $|\Psi_{\tilde{k}_1}^0\rangle$ of the
unperturbed case $\gamma=0$. Upon increase of $\gamma$ it becomes a new
eigenstate $|\Psi_{\tilde{k}_1}\rangle$, which will have overlap with several
eigenstates of the $\gamma=0$ case. We expand the eigenfunction of the
perturbed system to the first order approximation:
\begin{equation}
|\Psi_{\tilde{k}_1}\rangle=|\Psi_{\tilde{k}_1}^0\rangle+\gamma\sum_{{{k}_1^{\prime}}\neq
\tilde{k}_1}\frac{\langle\Psi_{k_1^{\prime}}^{0}|\hat{H}_1|\Psi_{\tilde{k}_1}^0\rangle}{{E_{\tilde{k}_1}^0}-E_{{k}_1^{\prime}}^0}|\Psi_{{k}_1^{\prime}}^0\rangle.
\end{equation}
The perturbation of strength $\gamma$ is local in the matrix representation
(\ref{matrix}), thus the relevant perturbation parameter is $\gamma/f$. This
has to be compared to the typical spacing of unperturbed eigenenergies.
For Bloch wave numbers far from $\pm \pi$ the spacing is of order $1/f$,
so the approximation should work for $\gamma < 1$.
For Bloch wave numbers close to $\pm \pi$ the approximation breaks down if
$\gamma \geq \pi -|k|$.

The off-diagonal ($k_1\neq\tilde{k}_1$) weight function at the
first order is given by :
\begin{equation}
C(k_1;\tilde{k}_1) \equiv  |\langle\Psi_{k_1}^0|\Psi_{\tilde{k}_1}\rangle|^2 =
\frac{|\langle\Psi_{k_1}^0|\hat{H}_1|\Psi_{\tilde{k}_1}^0\rangle|^2}{|E_{\tilde{k}_1}^0-E_{{k}_1}^0|^2}.
\end{equation}
$E_{k_1}^0$ and $E_{\tilde{k}_1}^0$ are the eigenenergies of the unperturbed
system. With $\Delta=k_1-\tilde{k}_1$ the weight function can be rewritten
in the following form 
\begin{widetext}
 \begin{equation}\label{eq:correlation}
C(k_1;\tilde{k}_1) = \frac{A^2\gamma^2}
{64(f+1)^2\cos^2(\frac{k}{2})\sin^2(\frac{\Delta}{2})\Big[\sin(\frac{2\tilde{k}_1+k}{2})\cos(\frac{\Delta}{2})
+ \cos(\frac{2\tilde{k}_1+k}{2})\sin(\frac{\Delta}{2})\Big]^2}, \; \;  \;
k_1\neq\tilde{k}_1,
\end{equation}
%\end{widetext}
%\clearpage
where $A$ is a constant. For $\gamma= 0$, $|\Psi_{\tilde{k}_1}\rangle =
|\Psi_{\tilde{k}_1}^0\rangle$, and the weight function is compact. For
$|\Delta|\ll 1$
%\begin{widetext}
\begin{equation}
C(k_1;\tilde{k}_1) \approx \frac{\gamma^2}
{(f+1)^2\frac{{\Delta}^2}{2}\cos^2(\frac{k}{2})\Big[\sin(\frac{2\tilde{k}_1+k}{2})
+\frac{\Delta}{2}\cos(\frac{2\tilde{k}_1+k}{2})\Big]^2}, \; \;  \;
k_1\neq\tilde{k}_1.
\end{equation}
\end{widetext}

From this formula we obtain several interesting results. First of all, the
decay of the weight function with increasing $\Delta$ means that we
have localization in normal mode space. For $2\tilde{k}_1 + k\neq 0,2\pi$, we
have
\begin{equation}
C\sim \frac{\gamma^2}{(f+1)^2}\frac{1}{\Delta^2}.
\end{equation}
We find algebraic decay $\sim 1/\Delta^2$ of the weight function, and for
$\gamma\to 0$ or $f\to\infty$ the weight function compactifies. For
$2\tilde{k}_1 + k = 0,2\pi$, we have
\begin{equation}
C\sim \frac{\gamma^2}{(f+1)^2}\frac{1}{\Delta^4}.
\end{equation}
Here we find algebraic decay $\sim 1/\Delta^4$ of the weight function,
that	
also compactifies when $\gamma\to 0$ or $f\to\infty$. Finally, for $k$ close to
$\pm\pi$
and large $f$, $C\sim \gamma^2/\Delta^2$. Thus we find that the $f$-dependence
drops
out for staggered states $|k\pm\pi|\ll 1$, and these states do not compactify
for $f\to\infty$. That is a remarkable quantum interference property, since
both simple intuition (see above) and classical theory predict the opposite.

\begin{figure}[h!]
\includegraphics[width=3.3in]{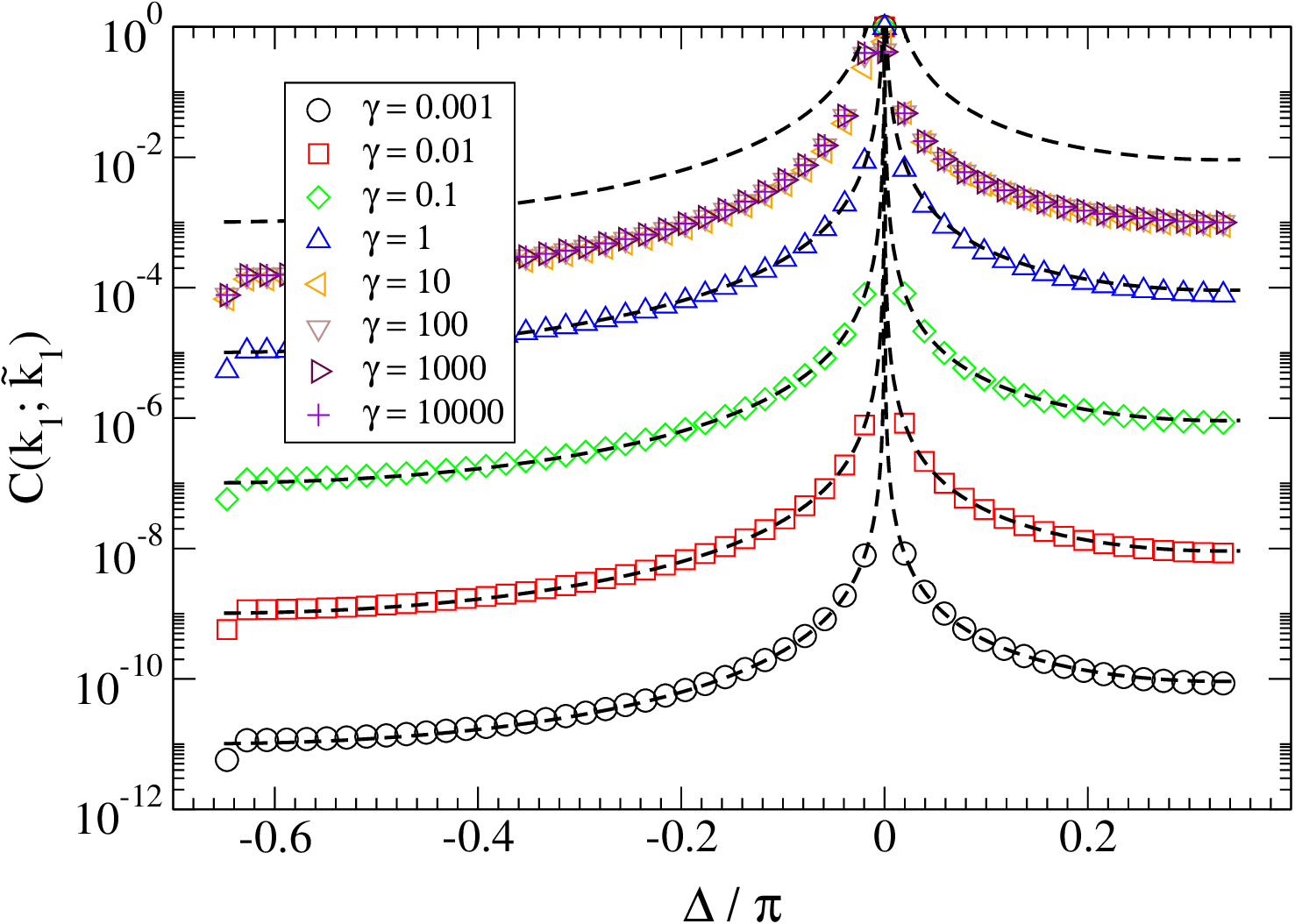}
\caption{\label{gcorr}Weight function for different values of the
interaction $\gamma$. Here f=101, $k=0$, and
$\tilde{k}_1=\frac{2}{3}\pi$. Dashed lines are results using the formula
(\ref{eq:correlation}) with $A^2=3.8$.}
\end{figure}

\begin{figure}[h!]
\includegraphics[width=3.3in]{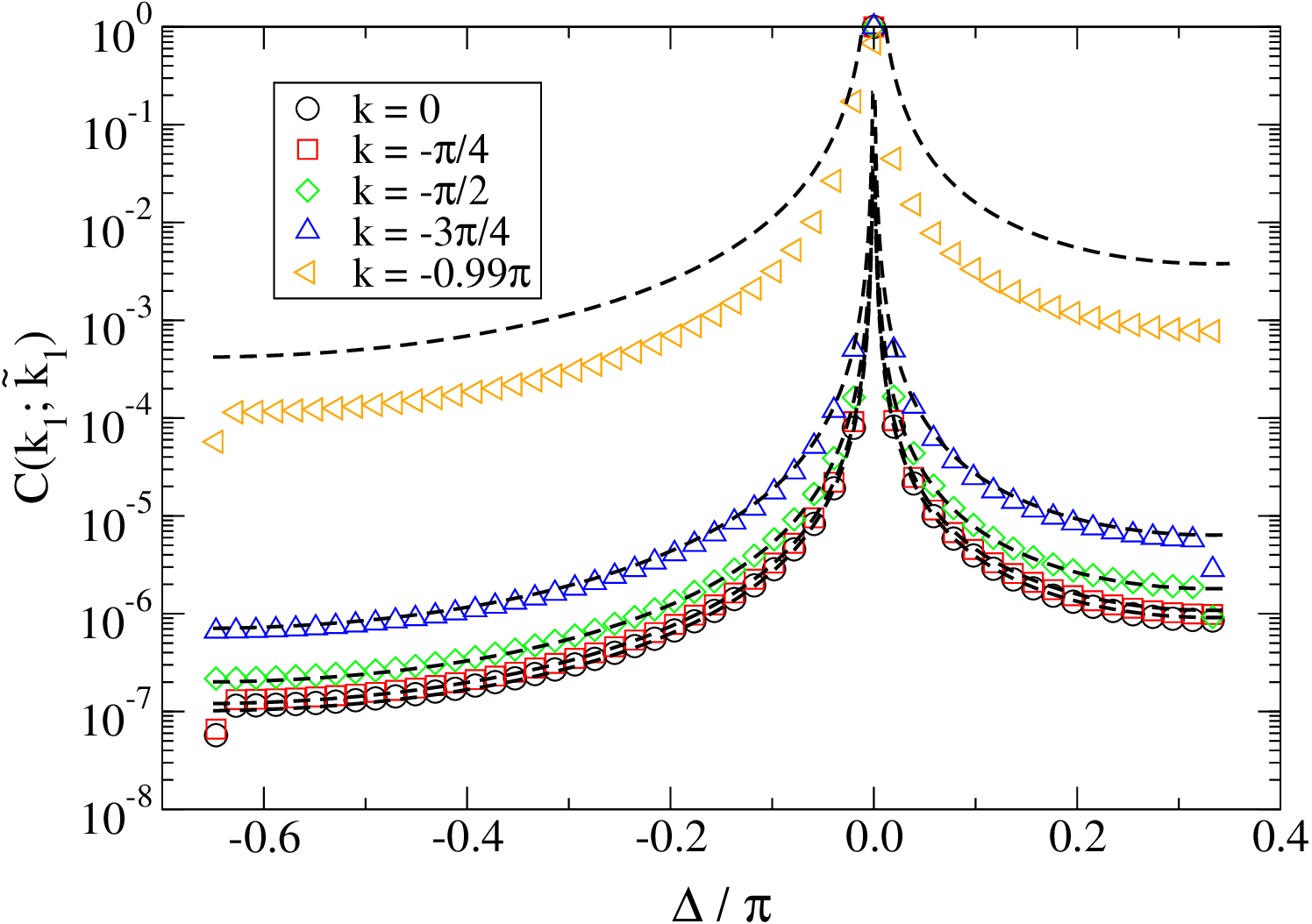}
\caption{\label{kcorr}Weight function for different values of the
Bloch wave number $k$. Here $\gamma=0.1$, $f=101$, and
$\tilde{k}_1+k/2=\frac{2}{3}\pi$. Dashed lines are results using the formula
(\ref{eq:correlation}) with $A^2=3.8$.}
\end{figure}

In Fig.\ref{gcorr} we show numerical results obtained by diagonalization of the
Hamiltonian for different values of $\gamma$.
%%%%%%%%%%%%%%%%%%%%%%%%%%%%%%%%%%%%%%%%%%%%%%%%%%%%%%%%%%%%%%%%%%%%%%%%%%%%%%%%%%
%%%%%%%%%%%%%%%%%%%%%%%%%%%%%%%%%%%%%%%%%%%%%%%%%%%%%%%%%%%%%%%%%%%%%%%%%%%%%%%%
We find localization in normal mode space, which can be interpreted as a
quantum q-breather. When increasing $\gamma$ the quantum q-breather becomes
less localized, and for large values of the interaction (from $\gamma=10$ on)
results do not change. The dashed lines are the results using the formula
(\ref{eq:correlation}) with $A^2=3.8$, value that was obtained by fitting the
numerical results for the lowest $\gamma$ ($=0.001$). We can see good agreement with
numerical results up to $\gamma=1$, beyond which perturbation theory does not
fit anymore.
In Fig.\ref{kcorr}, we show that the weight function is more localized
for $k=0$ and less localized for $k\to -\pi$. 
%Note that in the latter case
%perturbation theory does not fit the results. Indeed, in that
%limit the spacing between the unperturbed eigenenergies tends to zero,
%and higher orders in the perturbation theory have to be taken
%into account.
%the effective interaction increases up to the
%point that we can not use perturbation theory to describe the system {\bf SERGEJ: This
%is a naive statement I make based on what you commented to me some time ago about the
%interaction of energy levels as a function of $k$  when we go to $k=-\pi$ in
%the energy spectrum (see discussion on
%Fig.\ref{scorrind}). We can discuss}.
%%%%%%%%%%%%%%%%%%%%%%%%%%%%%%%%%%%%%%%%%%%%%%%%%%%%%%%%%%%%%%%%%%%%%%%%%%%%%
%%%%%%%%%%%%%%%%%%%%%%%%%%%%%%%%%%%%%%%%%%%%%%%%%%%%%%%%%%%%%%%%%%%%%%%%%%%%

\begin{figure}[h!]
\includegraphics[width=3.3in]{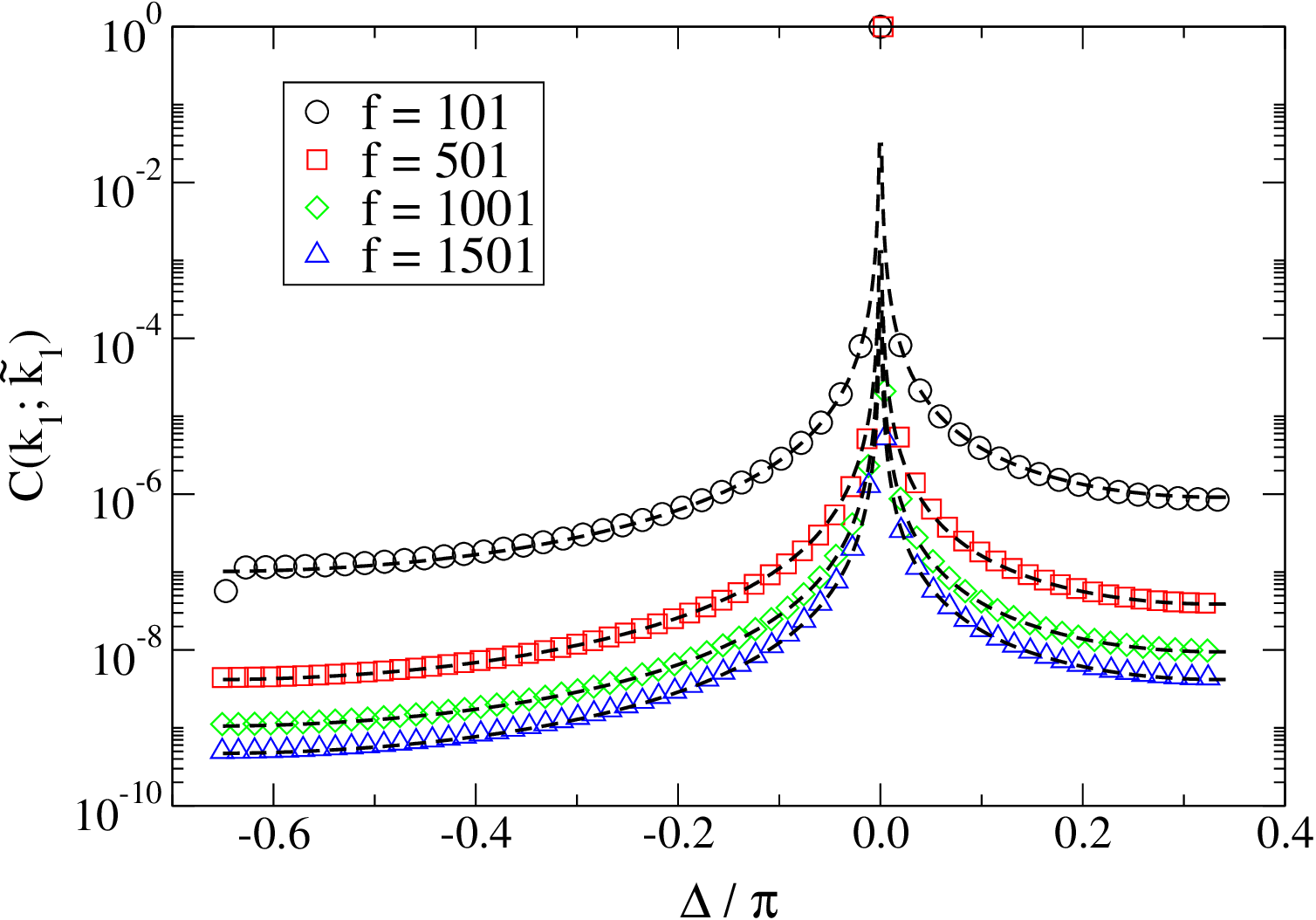}
\caption{\label{scorr}.Weight function for different sizes of the
system. Here $\gamma=0.1$, $k=0$, and $\tilde{k}_1\approx \frac{2}{3}\pi$ for
all curves. Dashed lines are results using the formula (\ref{eq:correlation})
with $A^2=3.8$.}
\end{figure}

\begin{figure}[h!]
\includegraphics[width=3.3in]{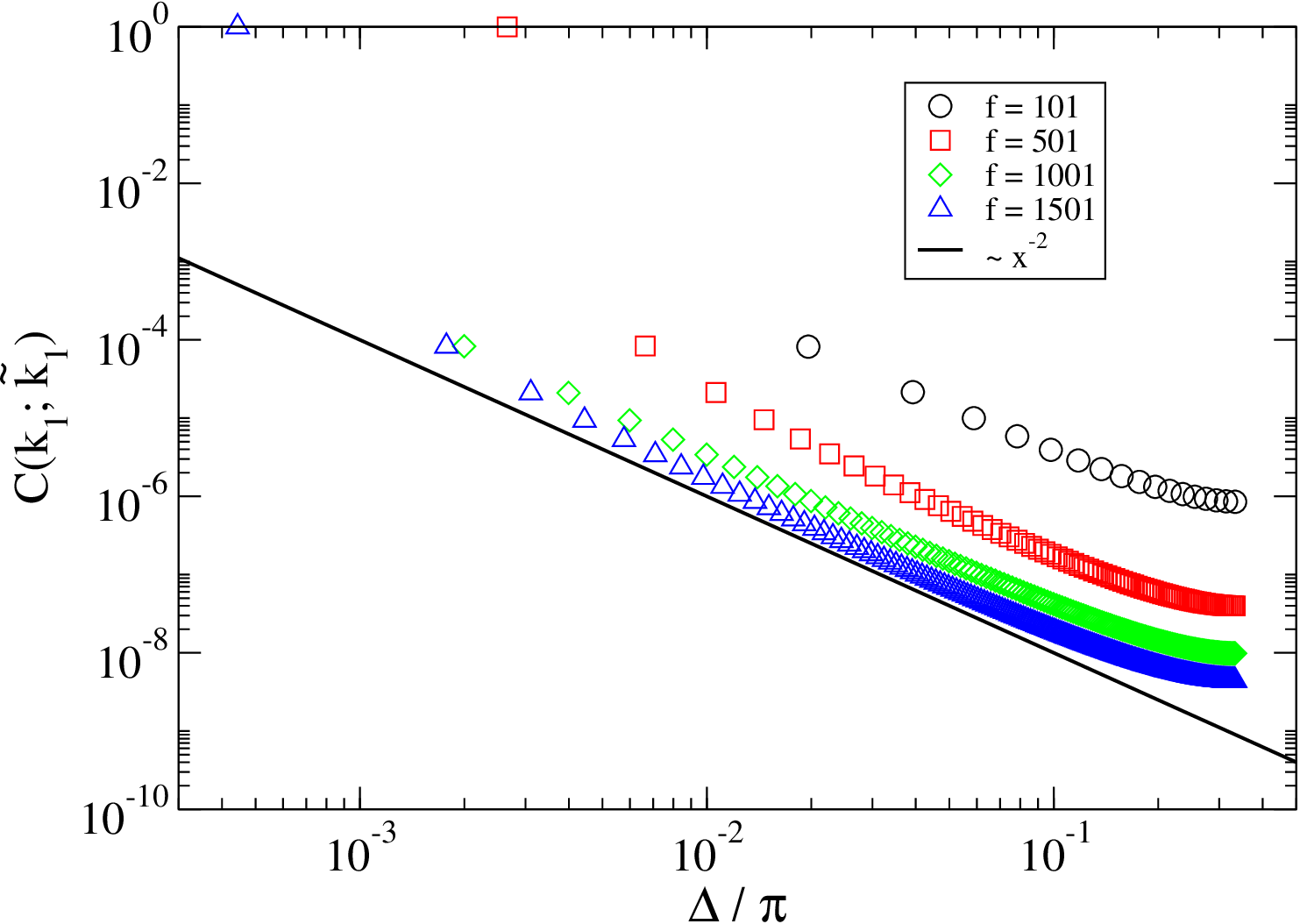}
\caption{\label{scorrlog}The same as in Fig.\ref{scorr} in log-log scale.}
\end{figure}

\begin{figure}[h!]
\includegraphics[width=3.3in]{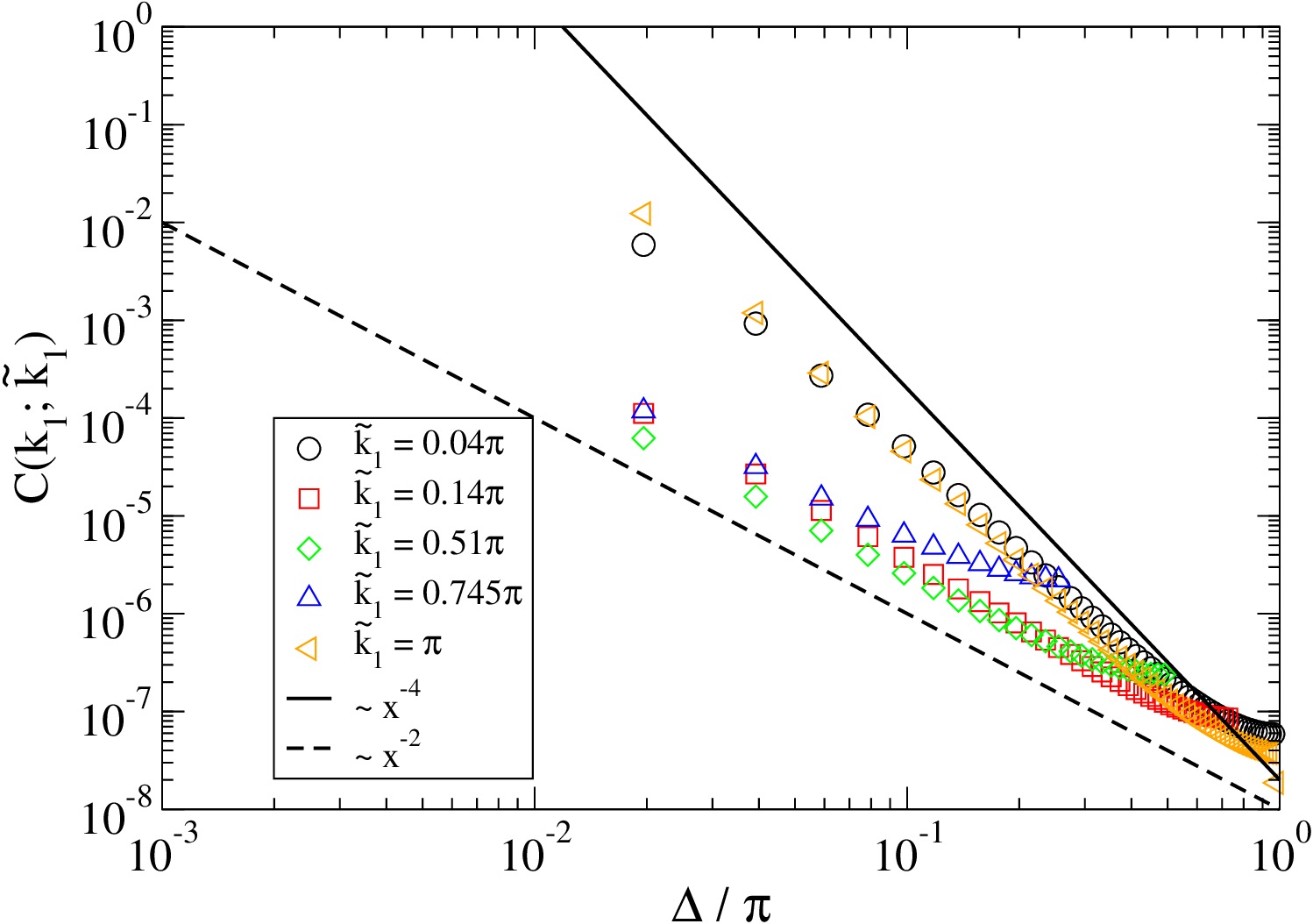}
\caption{\label{scorrlog2}Weight function for eigenstates
with different $\tilde{k}_1$. Here $\gamma=0.1$, $f=101$, and $k=0$.}
\end{figure}

While probing the influence of the size of the nonlinear quantum
lattice on the localization phenomenon, we find in Fig.\ref{scorr} that as the
size
increases the states compactify as we expected. In Fig.\ref{scorrlog} we see
the $1/\Delta^2$ decay for eigenstates fulfilling $2\tilde{k}_1+k\neq 0,2\pi$
($k=0$), and in Fig.\ref{scorrlog2} the $1/\Delta^4$ decay for
eigenstates fulfilling $2\tilde{k}_1+k=0,2\pi$ (also $k=0$). Both results agree
with the analytical results using perturbation theory.
%%%%%%%%%%%%%%%%%%%%%%%%%%%%%%%%%%%%%%%%%%%%%%%%%%%%%%%%%%%%%%%%%%%%%%%%%%%%%%%%
%----------------------------------------------------------------------------------------
%---------------------------------------------------------------------------------------
%%%%%%%%%%%%%%%%%%%%%%%%%%%%%%%%%%%%%%%%%%%%%%%%%%%%%%%%%%%%%%%%%%%%%%%%%%%%%%%%%%%%

\begin{figure}[h!]
\includegraphics[width=3.3in]{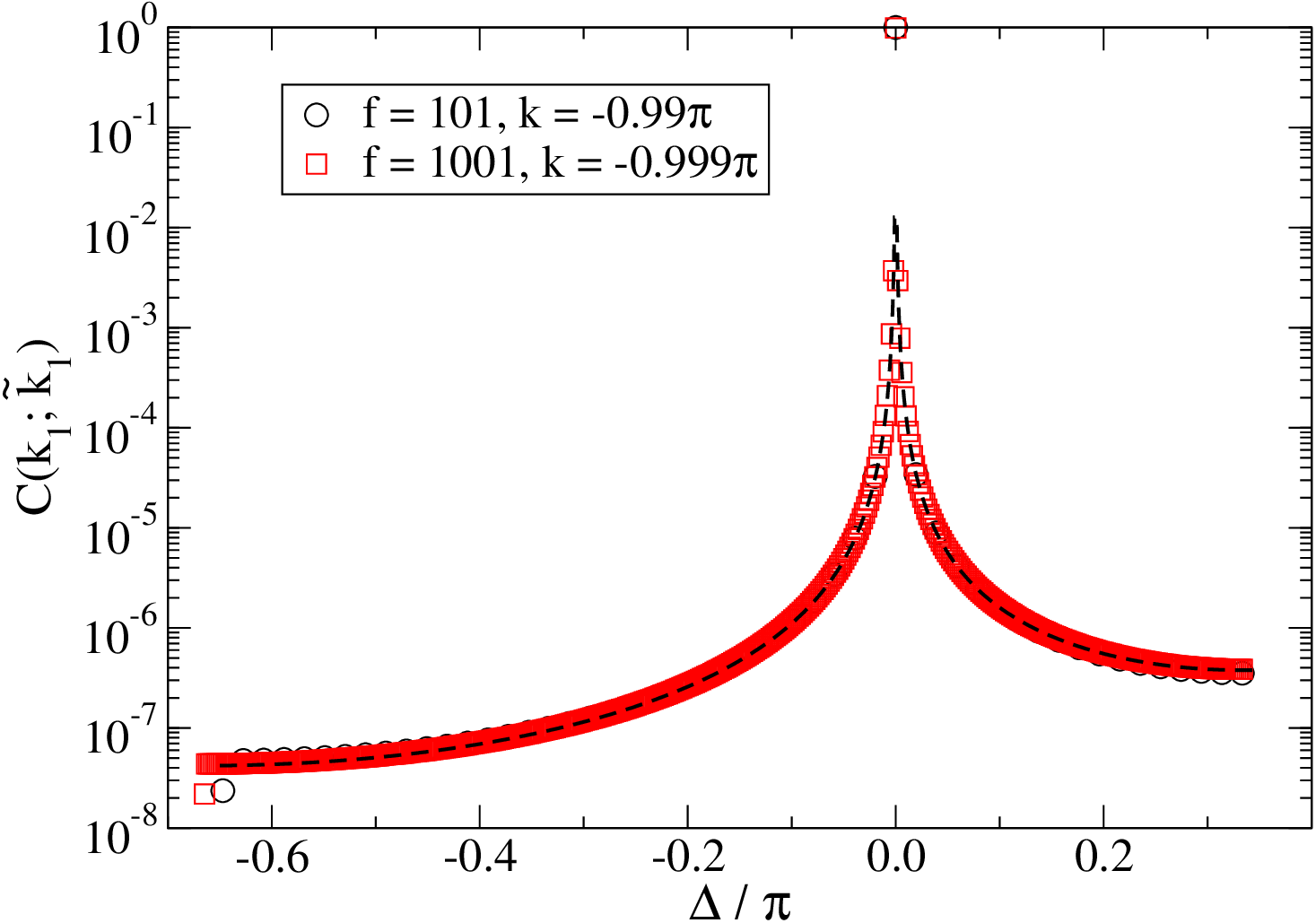}
\caption{\label{scorrind}Weight function for different sizes of the
system close to the band edge $k=-\pi$. Here $\gamma=0.001$ and
$\tilde{k}_1+k/2=\frac{2}{3}\pi$. The dashed line
is the result using the formula (\ref{eq:correlation}) with $A^2=3.8$.}
\end{figure}

In Fig.\ref{scorrind} we observe the predicted independence of the
localization phenomenon from the size of the system when $k$ is close to
$-\pi$. It is interesting that in this case the weight function does not
compactify in the dilute limit $f\to\infty$ as one would expect from simple
grounds.
The reason is that the larger $f$, the closer we can tune the Bloch
wave number to $\pm \pi$, where the perturbation expansion breaks down.
%%%%%%%%%%%%%%%%%%%%%%%%%%%%%%%%%%%%%%%%%%%%%%%%%%%%%%%%%%%%%%%%%%%%%%%%%%%%
%Note also that the formula (\ref{eq:correlation}) fits very well the
%numerical results although $k$ is close to
%$-\pi$. This is consistent with the argument presented above about the effective
%interaction when $k\to\pm\pi$. Since $\gamma$ is small the effective interaction
%when $k$ is close to $-\pi$ is still small enough to allow us to describe the
%system by perturbation theory and use the formula (\ref{eq:correlation}), in
%opposite to what happened in the case shown in Fig.\ref{kcorr}.
%%%%%%%%%%%%%%%%%%%%%%%%%%%%%%%%%%%%%%%%%%%%%%%%%%%%%%%%%%%%%%%%%%%%%%%%%%%%%%%%%%
%%%%%%%%%%%%%%%%%%%%%%%%%%%%%%%%%%%%%%%%%%%%%%%%%%%%%%%%%%%%%%%%%%%%%%%%%%%%%%%%%%

\section{Conclusions}

In this work we studied the properties of quantum q-breathers in a
one-dimensional chain containing two quanta modeled by the Bose-Hubbard
Hamiltonian. To explore localization phenomena in this system we computed
appropriate weight functions of the eigenstates in the normal mode
space using both perturbation theory and numerical diagonalization. We observe
localization of these weight functions, that is
interpreted as a signature of quantum q-breathers. The localization is
stronger when the size of the system increases. Unlike the classical case
where the localization is exponential, here we found
algebraic localization. This is a long range behavior, which follows
from the fact that the interaction $\gamma$ induces a linear perturbation
of the eigenstates which is local in real space, and also local
in the matrix representation in (\ref{matrix}).
That induces a mean-field type interaction between the normal modes, and
naturally
leads to algebraic localization. 
Note that the matrix (\ref{matrix}) is formally analogous to a semi-infinite
tight-binding chain with a defect at one end. Nevertheless it appears in our
context when starting with a translationally invariant system, but with
many-particle states which include interaction.

Since the effective interaction strength drops in the dilute
limit of large chains, we observe stronger localization (except for the
case of staggered states). The crucial difference to the classical
model is, that while the linear classical dynamics coincides with
the single particle quantum problem, nonlinearity in the classical model
effectively deforms the single particle dynamics (and adds many other features
like chaos etc). The interaction in the quantum problem takes the wave function
into
the new Hilbert space of many-body wave functions, which is still a linear
space, but higher dimensional.
Another feature of that quantum interaction is the fact that staggered states
do not compactify in the dilute limit of large chains. That property is
based on the interference of quantum states, and is not observed 
in the corresponding classical nonlinear equation.
A similar (yet weaker) signature of quantum interference is the observed
change of the power of the algebraic decay from two (generic)
to four when choosing particular
values of the wave number $k_1$, which depend on the Bloch
wave number $k$.
And yet another signature of quantum q-breathers is the fact that
they keep a finite localization in the limit $\gamma \rightarrow \infty$
as seen in Fig.\ref{gcorr}, and at variance to their classical counterparts,
which turn from exponentially localized to completely delocalized in that
limit.
The reason is that in this strong interaction limit extended two-boson states
correspond to their noninteracting counterparts which are projected onto the
basis space which does not contain doubly occupied chain sites, while strong
nonlinearity
in the classical problem completely deforms periodic orbits of the
noninteracting
system. 

We are aware of the fact that the quantum problem studied here is a linear one
(in terms of differential equations). Its correspondence to a classical nonlinear
system can be observed in the limit of many bosons
when treating the many particle 
quantum states within a Hartree approximation, which projects onto
product states. Often the classical description is also
achieved using suitable (e.g. coherent state) representations. The presented results
have an unambigous meaning in the chosen basis of the noninteracting system. Yet they will
of course in general depend on the chosen basis. Therefore it remains a
puzzling question, how to restore exponential localization of classical
q-breathers from the algebraic decay of quantum q-breathers with two bosons, in the limit
of larger numbers of bosons.
The fate of quantum q-breathers in higher dimensional
lattices is another interesting open question, which will be left
to future work. 
%\\
%\\
%\\
\acknowledgments
%\\
Dr Jean Pierre Nguenang acknowledges the warm hospitality of the Max
Planck Institute for the Physics of Complex Systems in Dresden. This work was
supported by the DFG (grant No. FL200/8) and by the ESF network-programme AQDJJ.

%\\

\end{document}